\documentclass[conference,a4paper]{IEEEtran}
\IEEEoverridecommandlockouts
\usepackage{cite}
\usepackage{amsmath,amssymb,amsfonts}
\usepackage{algorithmic}
\usepackage{graphicx}
\usepackage{textcomp}
\usepackage{xcolor}
\usepackage{xurl}
\usepackage[scaled]{helvet}

\def\BibTeX{{\rm B\kern-.05em{\sc i\kern-.025em b}\kern-.08em
    T\kern-.1667em\lower.7ex\hbox{E}\kern-.125emX}}

\usepackage{verbatim} 

\usepackage{graphicx} 
\usepackage{caption,subcaption}

\usepackage{array}
\usepackage{booktabs}
\usepackage{hyperref}

\begin{document}

\title{Adversarial Stress Testing of Outlier Detection in Subjective Image Quality Assessment\thanks{Funded by the Deutsche Forschungsgemeinschaft (DFG, German Research Foundation) -- DFG Project ID 251654672 -- TRR 161.}
}

\author{\IEEEauthorblockN{
First Author\IEEEauthorrefmark{1},
Other Authors\IEEEauthorrefmark{2}
}
\IEEEauthorblockA{\IEEEauthorrefmark{1}\textit{Institution First Author} }
\IEEEauthorblockA{\IEEEauthorrefmark{2}\textit{Institution Other Authors}} 
\IEEEoverridecommandlockouts
\IEEEpubid{\makebox[\columnwidth]{979-8-3503-1173-0/23/\$31.00
\copyright 2025 IEEE \hfill} \hspace{\columnsep}\makebox[\columnwidth]{ }}
}

\author{\IEEEauthorblockN{
Dietmar Saupe
}
\IEEEauthorblockA{\textit{Department of Computer and Information Science, University of Konstanz, Konstanz, Germany } } 
\IEEEoverridecommandlockouts
\IEEEpubid{\makebox[\columnwidth]{979-8-3503-1173-0/23/\$31.00
\copyright 2023 IEEE \hfill} \hspace{\columnsep}\makebox[\columnwidth]{ }}
}

\maketitle
\thispagestyle{plain}
\pagestyle{plain}

\begin{abstract}

In subjective image and video quality assessment, observers rate or compare selected stimuli. Before calculating mean opinion scores (MOSs), unreliable ratings should be identified and handled as outliers. Several outlier-detection methods are available, including standardized procedures, but their comparative performance is often evaluated using only specific types of synthetic outliers such as random clickers. Such tests do not necessarily reveal the worst-case behavior of these methods. To address this gap, we introduce and demonstrate a general empirical worst-case framework for outlier-detection methods, with proof-of-concept adversarial attack generators for both discrete absolute category and continuous visual analog scale ratings. The attacks use optimization algorithms to identify ratings that maximize the discrepancy between the resulting MOS estimates and the ground truth. We apply the proposed framework to several hard and soft outlier-detection methods and demonstrate substantial differences in their worst-case reconstruction performance under adversarial stress. We also propose several low-complexity outlier-detection methods that achieve excellent empirical worst-case performance. 
\end{abstract} 


\section{Introduction}
\label{sec_introduction}

Multimedia system developers and content providers strive to maximize the quality of experience (QoE) in human consumption of the digital media they produce. 
Therefore, many methods have been established and standardized to quantitatively assess the perceived quality of media.  Such assessments are typically carried out in special laboratories that provide controlled conditions during the tests or in crowdsourcing environments. Crowdsourcing has the advantage that it is easier to recruit many participants of specific target groups such as from a certain country or age group, while it is harder to control for the viewing conditions that typically differ between participants. 

The recruited observers provide the quality values directly, for example, on the discrete 5-level absolute category rating scale (ACR), on a continuous visual analog scale (VAS), or indirectly by comparing perceived quality between two or more presented stimuli. From the raw data collected, the experimenter then reconstructs the desired stimulus quality on a linear scale as perceived by the group of observers. For the ACR and VAS data corresponding to a particular image, this can simply be the corresponding mean opinion score (MOS), which is the average of the discrete resp.\ continuous ratings. Recently, better statistical methods that account for subject bias, inconsistency and more have been established and standardized (see Section \ref{subsec_Soft_outlier_detection}). For the indirect quality assessment by pairwise comparisons, a more complex reconstruction algorithm must be applied. A typical choice to convert qualitative pairwise comparison data into a quantitative interval scale is maximum likelihood estimation of the parameters of a probabilistic model based on Thurstone’s law of comparative judgment. 

Reliable subjective image quality assessment hinges on attentive and well-informed participants. However, inattention, fatigue, personal biases, strategic behavior, technical issues, and deviations from prescribed viewing conditions can all lead to inconsistent or unreliable ratings.  Furthermore, data quality can be compromised by spammers, data entry errors, and outlier responses. Addressing these potential issues through careful experimental design, robust software, clear instructions, and participant screening or data validation steps are crucial to obtain reliable and meaningful results in subjective visual quality assessments.  For crowdsourcing, techniques for identifying unreliable observers are particularly important.

\subsection{Performance evaluation of outlier-detection methods}
\label{SubSecRMSD}

We use the term \textit{outlier detection} jointly for methods that detect and handle such unreliable observers and outliers. The handling typically consists of either discarding the data of unreliable observers and outliers (hard outlier detection) or by weighting them according to the degree of their reliability (soft outlier detection). Soft outlier-detection methods are integral parts of the scale reconstruction process.

Generally, outlier detection identifies observers whose data significantly deviate from the consensus of the data provided by the reliable majority of observers. Several outlier-detection methods for QoE studies have been introduced in the literature and some have been standardized by the International Telecommunication Union (ITU). Therefore, there is a need to measure and compare the performance of these and newly proposed outlier-detection methods. 

Recently, the performance of several quality-reconstruction methods, including outlier-detection methods, has been evaluated by comparing their outputs on clean and synthetically corrupted datasets, generated, for example, by randomly changing some observer ratings or by adding some spammers; see Section \ref{subsec_Analysis_of_robustness}. 
Each method was applied to both datasets and the resulting estimations were compared using the root mean square difference (RMSD) to assess the impact of noise. A method with a smaller difference between the reconstructions before and after noise injection was considered less sensitive to that particular corruption procedure.

\subsection{Limitations of current approaches}

The above approach seems natural and generally accepted. However, it has two limitations that we address in our contribution. The first limitation is that the resulting comparison depends strongly on the chosen synthetic observers. The second is that the difference between reconstructions before and after corruption does not necessarily measure the accuracy of the reconstructed quality values with respect to the true quality. We begin by reviewing two examples showing that synthetic observers can yield inconsequential results.

In Altieri et al.\  \cite{altieri2024subjective} (Figs.\ 4, 5), three soft outlier handling methods, ESQR, ZREC, and SUREAL, were compared on two datasets, one of which was KonIQ-10k \cite{hosu2020koniq}, using two types of artificial observers. The first type was noise injection in which a given fraction of opinion scores was replaced by random scores from 1 to 5. For the second type, a given number of spammers were added who randomly annotated all stimuli from 1 to 5. The RMSD between the reconstructed scale values with the soft outlier handling techniques before and after the introduction of random annotations was reported. For noise injection, ESQR was better than ZREC, providing a lower RMSD. However, for spammers, it was the opposite.

Similarly, in Li and Bampis \cite{li2017recover}, the outcome of the analysis was ambiguous. For the case of subject corruption, where all scores corresponding to a number of subjects were scrambled, their proposed method clearly outperformed the compared methods. In contrast, for the case of random corruption, where a percentage of raw scores was replaced by a random score of 1 to 5, it did not provide any advantage.

\begin{figure}[t]
    \centering
    \includegraphics[width=1\columnwidth]{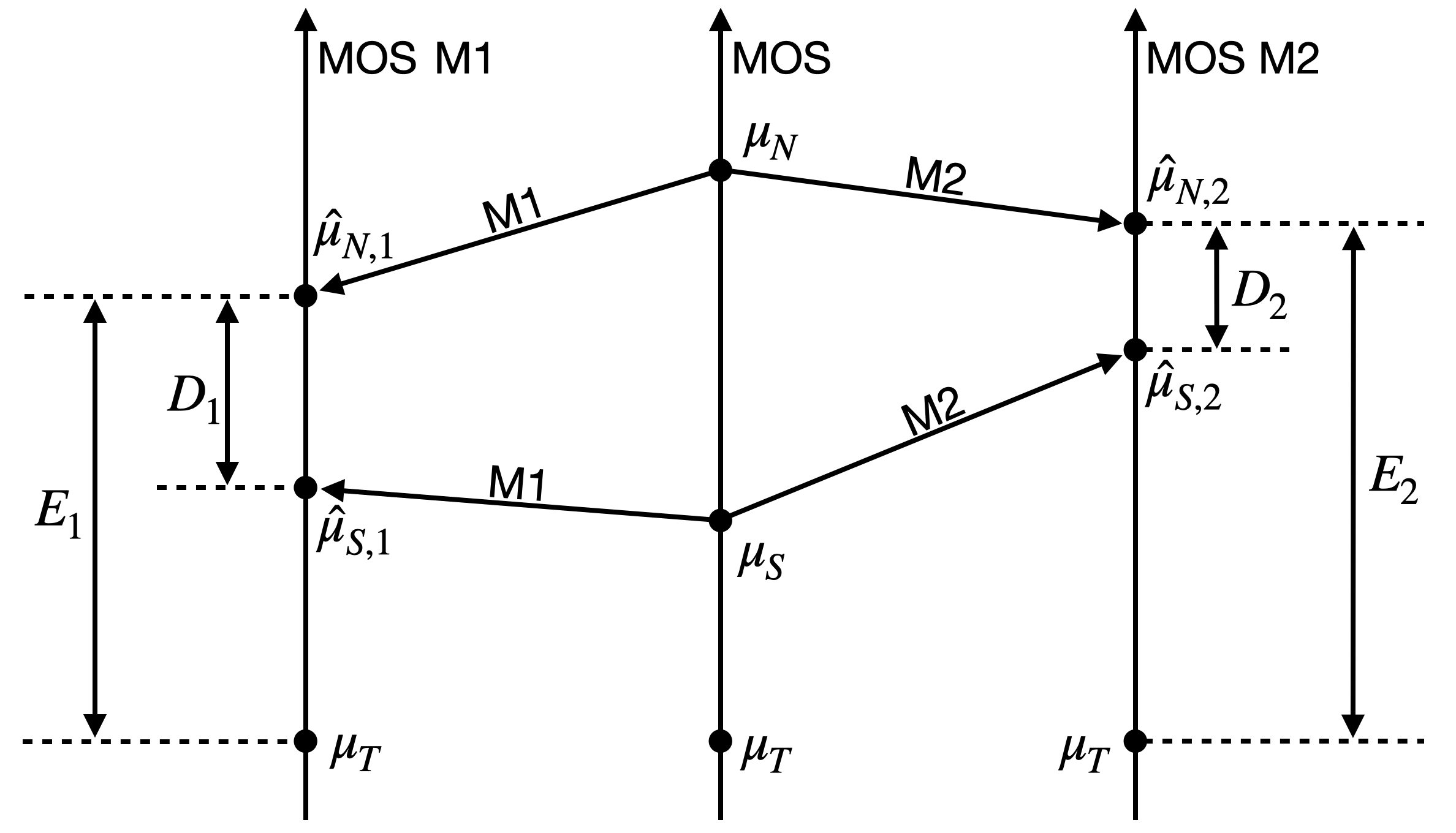}
    \caption{The MOS for one stimulus in the ground truth ($\mu_T$), the dataset sample ($\mu_S$) and the sample with injected noise ($\mu_N$). Two hard outlier detection methods (M1, M2) are compared, giving reconstructed MOS $\hat{\mu}_S$ and $\hat{\mu}_N$. Method M2 is less sensitive to this particular noise injection, since $D_2<D_1$, but Method M1 has the smaller reconstruction error relative to the ground truth, since $E_1<E_2$.}
    \label{fig_AB}
\end{figure}

Since it is unknown which types of outliers can be expected in a quality assessment study, it would be desirable to have an outlier detection method which generally works the best, even in the worst case. However, the behavior of outliers can be very diverse; they may follow a structured pattern, act adversarially, or exhibit entirely different characteristics. Surely, it is not feasible to create and analyze a set of scenarios that capture all possible outlier behaviors. 

The second limitation is that the RMSD robustness measure, based on the difference between reconstructions from data with and without synthetic noise, may also be inconclusive.  Quality reconstruction methods aim to estimate the true quality of stimuli based on a sample of human ratings.  However, when comparing reconstructions from noisy and clean rating samples, processed by different outlier-detection methods, the choice of the more robust method that has a smaller RMSD does not necessarily produce a result with a smaller error w.r.t.\ the true quality. 

This situation is illustrated in Figure \ref{fig_AB} in which we compare two hard outlier-detection methods (M1, M2) and use the MOS for scale reconstruction. 
M1 and M2 are applied to the original and noisy dataset sample with results shown on the left and right scales, respectively. The scale increase due to noise injection is smaller for Method M2 ($D_2 < D_1$), so Method M2 appears to provide the more robust outlier detection. However, its reconstruction error is larger than that of Method M1, which indicates that M1 should be favored over M2. 

Proponents of the RMSD criterion might argue that in their approach \textit{clean} datasets $S$ were used that can be trusted to faithfully represent the ground truth, so that the error in the MOS ${\mu}_{N}$, derived from noisy sample data, can be estimated as RMSD w.r.t.\ $\hat{\mu}_{S}$. However, this estimate is weak, as other scale reconstruction methods with outlier detection will typically generate a different and contradictory representative $\hat{\mu}_S$ of the ground truth.

Finally, we highlight that the above limitation is not artificial or of little significance but can actually be observed in practice having a very large effect. A striking example is the ITU standardized outlier-detection method LPCC, compared to six other established methods for the case of added spammers. 
In our analysis, LPCC clearly appeared to perform best according to the RMSD criterion for this particular spammer experiment, but its reconstruction error was about twice as large as that of the other methods. See Table \ref{tab_spammers} in Section \ref{sec_results} for more details.

\subsection{Evaluation of outlier-detection performance by RMSE in the worst-case scenario by adversarial attacks}

We have shown that the robustness measured as the RMSD between the scales derived from a sample dataset with and without injection of noise or spammers is not conclusive about the performance of an outlier-detection method. Instead, it is necessary to assess the root mean square error (RMSE) w.r.t.\ ground truth of the perceived quality. This ground truth can be given in the form of an appropriate and accepted statistical model of observer ratings. This has another advantage, namely that with such models any number of sample datasets can be drawn and analyzed. This facilitates further useful statistical analysis, such as significance tests and effect size estimation.

In addition, we adopt an empirical worst-case analysis approach. Rather than relying on hand-crafted models to identify outliers, such as spammers, we utilize adversarial attacks to assess the vulnerabilities of outlier removal methods. An adversarial attack introduces ratings from one or more malicious subjects with the goal of distorting the quality estimates produced by the assessment system. The attacker designs these ratings to exploit weaknesses in the outlier-detection and quality-reconstruction procedures, thereby producing a structured, coordinated attack rather than independent random responses. Consequently, adversarial attacks serve as generally applicable stress tests, allowing us to evaluate the effectiveness of various methods in handling worst-case scenarios.

The optimization of an attack depends on the type of rating data. For discrete data, such as absolute category ratings or pairwise comparisons, the attacker’s responses are selected from a finite set of alternatives. For example, five-level ACR responses are restricted to integers $(1,\ldots,5)$, while pairwise-comparison responses are restricted to “left” or “right.” Attack design in this setting is therefore a combinatorial optimization problem.

In contrast, when ratings are collected on an interval scale, the attacker’s responses are continuous variables. This allows the attack to be formulated as a continuous constrained optimization problem. Gradient-based solvers can then be applied when the processing pipeline is differentiable or sufficiently smooth for numerical gradient estimation.

Through this approach, various outlier-detection methods are evaluated according to their empirical worst-case reconstruction performance under adversarial attacks. The success of an attack is quantified by measuring the accuracy of the reconstructed scale values using the RMSE in relation to the ground truth. Each attack scenario is executed 250 times across different datasets sampled from the ground truth models, ensuring that the results provide a comprehensive assessment that captures general trends, rather than being influenced by the characteristics of a single dataset.

Our findings facilitate a comparative analysis of different outlier-detection methods, focusing on their worst-case performance. The results yield valuable insights into how effectively each method mitigates the impact of malicious manipulations, thereby improving our understanding of their efficacy in subjective quality assessment contexts. 

The main contributions of this work are:
\begin{itemize}
\item A framework for evaluating outlier-detection methods in subjective quality assessment by means of adversarial attacks and empirical worst-case analysis.
\item Attack-generation procedures for discrete ACR data and continuous VAS data, using a genetic algorithm and a constrained numerical optimizer, respectively.
\item An analysis of why comparisons based only on clean-versus-corrupted reconstruction differences can be inconclusive.
\item New low-complexity hard and soft outlier-detection methods used to demonstrate the application of the framework.
\item An empirical comparison of established and proposed methods under the resulting adversarial stress tests.
\end{itemize}

This paper extends our previous work on adversarial attacks for discrete ACR-type ratings \cite{saupe2025robustness} by formulating a corresponding attack for continuous VAS ratings. Taken together, the two cases provide proof-of-concept implementations of a general empirical worst-case framework: a genetic algorithm is used for the discrete ACR setting, while a bounded numerical optimizer is used for selected smooth reconstruction pipelines in the continuous VAS setting. Neither implementation is intended to establish the globally optimal attack for its respective search space. Rather, the experiments demonstrate how adversarial attack generation can be used as a practical methodology to expose and compare failure modes of outlier-detection methods.

\section{outlier-detection methods and related work}
\label{sec_related_work}

In this section, we extract from the literature the outlier handling methods in subjective quality assessment by the single stimulus method on a visual analog scale or from absolute category ratings. We also propose two new hard and soft outlier-detection methods. Most of these methods have been implemented and evaluated in our study, in part making use of code provided by \cite{altieri2024subjective}. Due to space limitations, we cannot fully describe the algorithms for combined scale reconstruction and outlier handling and refer the reader to the given references. In addition, we provide a MATLAB implementation of the methods together with adversarial attacks. 

 \subsection{Hard outlier-detection methods}

There are ITU standards for subjective evaluation of audiovisual quality in multimedia. At least two of them, ITU-R BT.500-15 \cite{itu_bt500} and ITU-T P.910 \cite{itu_p910} include three recommendations on how to screen observers. However, the ITU has not evaluated these methods with respect to robustness.
\smallskip

\noindent \textbf{KB.} The kurtosis-based method (KB) in ITU-R BT.500 \cite{itu_bt500} is a hard outlier method that detects outliers by evaluating how many times a subject voted far above and below the MOS. Two conditions are computed for each subject. The first is the overall percentage of these extreme ratings and the second sets the counts in relation. If both conditions exceed a  threshold, the subject is removed.
\smallskip

\noindent \textbf{CB.} ITU-R BT.500 \cite{itu_bt500} specifies another hard outlier detection that is correlation-based (CB). The minimum of linear Pearson and rank-order Spearman correlation is calculated between each observer's scores and the MOS of the items. An acceptance threshold is determined on the basis of the mean and variance of these values of all observers. Subjects with a minimum correlation below this threshold are removed. 
\smallskip

\noindent \textbf{LPCC.} The outlier-detection method in ITU-T P.910 \cite{itu_p910} iteratively removes observers based on Linear Pearson Correlation Coefficients  (LPCC). In each iteration, the MOS is recalculated for the current inlier set, and the LPCC of each subject's ratings is computed with this updated MOS. The inlier subject with the lowest correlation is removed if LPCC $< 0.75$.
\smallskip

\noindent \textbf{HB.} We include an entropy-based outlier-detection method (HB) proposed by He et al.\ in \cite{he2005optimization} that has not yet been applied to datasets for visual quality assessment. The entropy of the probability distribution of ARC-type ratings of a stimulus is maximal for the uniform distribution and minimal when all ACR ratings are equal. Therefore, this approach assumes that observer consensus is maximized when the total entropy of the ACR rating distributions (across all stimuli) is minimal. Using a greedy search algorithm, the method selects a specified number of outliers to minimize this total entropy.
\smallskip

We propose two additional hard outlier-detection methods, thresholding the mean absolute Z-score (MAZ) and the negative log-likelihood (NLL).
\smallskip

\noindent \textbf{MAZ.} Our first outlier-detection method is similar in spirit to the KB kurtosis-based approach in ITU-R BT.500. For each of the stimuli, the ratings are converted to z-scores and their absolute values are recorded. Observers with ratings having a mean absolute z-score greater than $1$ are discarded as outliers.  
\smallskip

\noindent \textbf{NLL.} Our second method is iterative. In each iteration, the observer is removed as an outlier whose ratings have the maximum negative log-likelihood (NLL) with respect to the rating distributions belonging to the current set of inliers. Let $p_j  = (p_j{(1)}, \ldots, p_j{(5)})$ denote the empirical probability distribution for the ACR ratings of the $j$-th stimulus, $j=1,\ldots,J$. Then the NLL for the set of all ratings $r_{i,j}$ of an observer $i$ (who belongs to the current inlier set) is given by $- \sum_{j = 1}^{J}\log p_j{(r_{i,j})}$. The iteration is terminated when all inlier observers have a NLL not exceeding a given threshold. For our implementation, we set the threshold to 1.31.

\subsection{Soft outlier-detection methods}
\label{subsec_Soft_outlier_detection}

Soft outlier-detection methods are integrated in the scale reconstruction process, which is based on statistical models of perceived stimulus quality. The ratings of subjects are considered as samples of Gaussian random variables that can incorporate parameters for the perceived quality of all stimuli and their dispersion as well as bias and inconsistency for all observers, as introduced in \cite{janowski2015accuracy}. 
\smallskip

\noindent \textbf{SUREAL.} Li et al. \cite{li2020simple} propose a simple model to estimate the true visual quality of a stimulus from subjective ratings. This model accounts for subject bias, inconsistency, and performs soft outlier rejection by down-weighting inconsistent raters.  The model's parameters are estimated using maximum likelihood estimation (MLE) and an Alternating Projection solver. It has been incorporated into ITU-T P.913 \cite{itu_p913} and ITU-R BT.500-15 \cite{itu_bt500}.
\smallskip

\noindent \textbf{ESQR.} Altieri et al.\ \cite{altieri2024subjective} proposed ESQR (Entropy-based Subjective Quality Recovery), a nonparametric algorithm that estimates subjective quality by weighting individual opinion scores based on their reliability.  Reliability is defined as the inverse of the information-theoretic surprise of an opinion score given the overall distribution of scores for that stimulus.  
This approach avoids the assumptions underlying parametric methods, leading to greater robustness and accuracy.
\smallskip

\noindent \textbf{RMLE.} Tiotsop et al.\ \cite{tiotsop2022regularized} introduced RMLE (Regularized Maximum Likelihood Estimation), a method for estimating subjective quality from noisy individual ratings, employing an MLE framework with a regularization term to reduce the influence of outlier scores.  
In contrast to ESQR's non-parametric approach, RMLE modifies the standard MLE framework to achieve robustness.
\smallskip

\noindent \textbf{ZREC.} Zhu et al.\ \cite{zhu2023ZREC} proposed ZREC (Z-score RECovery), a method to recover MOS from subjective quality scores.  Estimate subject bias, inconsistency, and content ambiguity using Z-scores, then adjust subject coefficients in the MOS calculations rather than completely rejecting outlier subjects. This approach was shown to be highly correlated with the more complex SUREAL method and provides MOS estimates with tight confidence intervals.
\smallskip

\noindent \textbf{DCC.} Previously, Li et al.\ introduced a similar soft method in \cite{li2017recover} at the Data Compression Conference (DCC, 2017), using the same model parameters, however, estimated by MLE in an iterative procedure. This approach was reported to sometimes suffer from convergence problems \cite{li2020simple}, and we did not use it in our study. 
\smallskip

We also propose modifications of two simple hard outlier-detection methods in order to turn them into soft outlier detection. The first is MAZ, based on Z-scored differences to MOS, and the other is the correlation-based method CB.
\smallskip

\noindent \textbf{MAZsoft.} As for MAZ, the ratings are converted to z-scores for each stimulus and the mean absolute z-score $z^a_{i}$ is calculated for each subject $i$. Reconstruction of image quality proceeds by weighted mean opinion scores with normalized weights $e^{-\alpha z^a_{i}}/\sum_k e^{-\alpha z^a_{k}}$ for the scores of subject $i$. The parameter $\alpha > 0$ determines the strength of the penalty for ratings from subjects with a large mean absolute z-score. The default is $\alpha = 1$. 
\smallskip

\noindent \textbf{CBsoft.} Only the Pearson correlation $r_i \in [-1,1]$ is calculated between the $i$-th observer's scores and the MOS of the items. The MOS is recomputed with normalized weights $e^{-\alpha r_{i}}/\sum_k e^{-\alpha (1-r_{k})}$ for the scores of subject $i$. The parameter $\alpha > 0$ determines the strength of the penalty for ratings from subjects with small Pearson correlations $r_i < 1$. The default is $\alpha = 1$. The rank-order Spearman correlation is a component of CB but is not included here to facilitate continuous optimization methods for adversarial attacks.

\subsection{Analysis of robustness}
\label{subsec_Analysis_of_robustness}

Analyses of the robustness of outlier detection in visual quality assessment were presented for the DCC, SUREAL, RMLE, and ESQR methods. In all four cases, these proceeded in the way as explained in the Introduction. Noisy and nonauthentic quality ratings were introduced for selected datasets and RMSD between the reconstructed scale values before and after injection of synthetic data was taken as an indicator of robustness of the methods. Simulations with the following types of noise have been carried out.

\begin{itemize}
    \item Scores of a number of subjects are scrambled (DCC \cite{li2017recover}, SUREAL \cite{li2020simple}).
    \item A percentage of the scores for each subject are scrambled (RMLE \cite{tiotsop2022regularized}).
    \item A raw score is replaced by a random score from 1 to 5 with a probability (DCC \cite{li2017recover}).
    \item For all or a number of subjects, a percentage of their ratings was changed into random numbers between 1 and 5, (ESQR \cite{altieri2024subjective} and \cite{tiotsop2023scoring}).
    \item Spammers, who score the quality at random, are added to the dataset (ESQR \cite{altieri2024subjective}).
\end{itemize}

Similarly, the influence of outliers in quality assessment using pair comparisons was studied in \cite{pastor2023recovering}. 

Studies evaluating the robustness of quality assessment methods against synthesized noise of the above types overlook the potential effects of other types of outliers and the impact of targeted attacks.  Adversarial attacks, unlike random noise, exploit algorithm weaknesses to cause significantly greater distortions, a vulnerability already explored in objective quality assessment \cite{NRIQA_attack}. These attacks involve imperceptible manipulations of the input image, leading to misleading quality predictions even without visible changes. However, for subjective quality assessment the effect of adversarial attacks on outlier detection and subsequent scale reconstruction has not yet been studied, a gap that is addressed in this contribution. 

\section{Adversarial worst-case framework}
\label{sec:adversarial_attacks}

\begin{figure}[t]
    \centering
    \includegraphics[width=\linewidth]{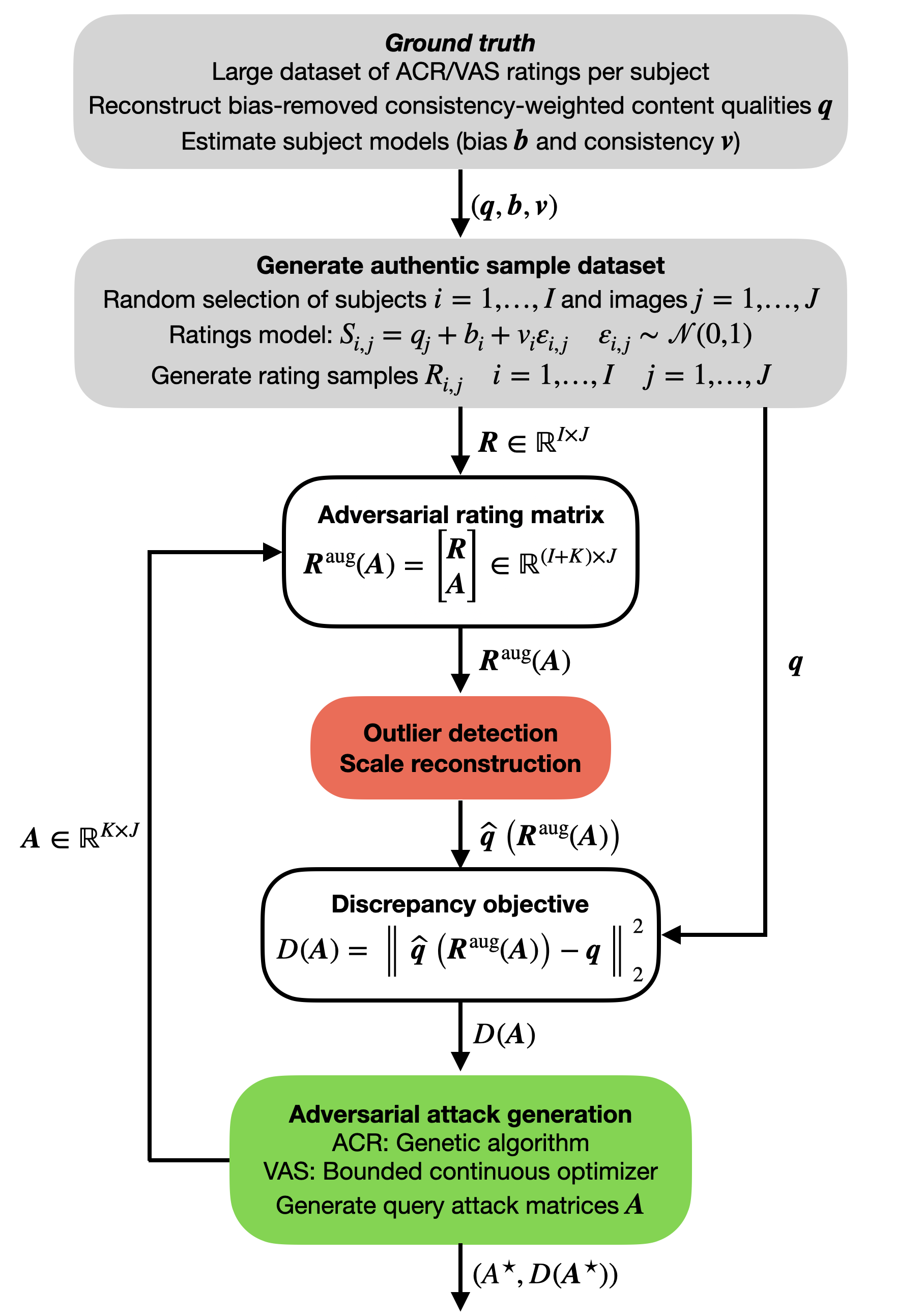}
    \caption{Overview of the proposed approach for estimating the worst case performance of an outlier detection and reconstruction method.}
    \label{fig_overview}
\end{figure}

We investigate the use of adversarial rating attacks as an empirical worst-case analysis tool for subjective image-quality assessment with outlier detection. The present study is restricted to two single-stimulus rating protocols: Absolute Category Rating and Visual Analog Scale. The framework is intended to separate the definition of the evaluation criterion — reconstruction error relative to ground truth — from the algorithm used to search for damaging rating configurations. Figure \ref{fig_overview} provides an overview. The framework has two separable components. First, an evaluation criterion is defined by the reconstruction error relative to an independently specified ground truth. Second, an attack-generation algorithm searches for an injected rating configuration that increases this error. The present paper instantiates the second component with a genetic algorithm for discrete ACR ratings and a bounded local numerical optimizer for continuous VAS ratings. Other attack-generation algorithms can be substituted without changing the evaluation principle.

\subsection{Ground truth and simulation of subjective ratings}
\label{subsec:rating_simulation}

To generate realistic and reproducible rating data, we use empirical
ratings in the cleansed KonIQ-10k dataset \cite{hosu2020koniq}, containing $10{,}073$ images rated
by $1{,}257$ subjects. We apply the SUREAL method to derive the parameters
required for the simulation, including ground truth quality values,
subject-dependent bias, and inconsistency parameters. Owing to the
preceding data-cleansing process and SUREAL's soft outlier-handling
procedure, the resulting subject models can reasonably be assumed to be
realistic and reliable. Subsequently, these models are used to generate
synthetic rating matrices while preserving the main statistical
characteristics of subjective empirical ratings.

For each simulated experiment, we randomly choose a set of $J$ images and
$I$ subjects. Each image $j=1,\ldots,J$ is assigned the corrresponding ground truth quality
value $q_j$. The ratings provided by subject $i=1,\ldots,I$ for image $j$ are
generated according to random variables
\begin{equation}
    S_{i,j}
    =
    q_j + b_i + v_i\epsilon_{i,j},
    \label{eq:rating_model}
\end{equation}
where $b_i$ denotes the bias of the subject $i$, $v_i$ denotes the inconsistency of the subject, and $\epsilon_{i,j}$ are independent standard Gaussian random variables,
\begin{equation}
    \epsilon_{i,j}\sim\mathcal{N}(0,1).
    \label{eq:subject_noise}
\end{equation}
Thus, the subject-specific random component has zero mean and a standard
deviation determined by $v_i$. This procedure allows the number of subjects,
images, and ratings to be independently controlled.

For the ACR condition, ratings are represented on the conventional five-point
scale. Discrete ACR ratings are obtained by rounding the continuous samples to
the nearest integer and clipping the result to the admissible five-point
scale,
\begin{equation}
    R_{i,j}
    =
    \min\!\left\{
        5,
        \max\!\left\{
            1,\operatorname{round}(S_{i,j})
        \right\}
    \right\}.
    \label{eq:acr_generation}
\end{equation}

For the VAS condition, the samples generated by the SUREAL model are retained
as continuous values. For consistency with the model, they are not clipped during the generation of the
reference ratings. Only a negligible fraction of the samples falls outside
the nominal interval $[1,5]$. However, during adversarial optimization, the
 ratings of attacker subjects $k = 1,\ldots,K$ are constrained to this interval to represent
feasible ratings from real subjects,
\begin{equation}
    1 \leq A_{k,j} \leq 5.
    \label{eq:vas_constraints}
\end{equation}

\subsection{Adversarial model}
\label{subsec:adversarial_model}

We consider an adversary whose objective is to impair the scale-reconstruction
pipeline by increasing the discrepancy between the reconstructed quality
values and the ground truth quality values. Let
$\widehat{\boldsymbol q}(\boldsymbol R)$ denote the quality vector
reconstructed from a rating matrix $\boldsymbol R$ that may include injected adversary ratings, and let
$\boldsymbol q$ denote the corresponding ground truth quality vector. We
measure the error of the quality reconstruction vector using the squared Euclidean discrepancy,
\begin{equation}
    \mathcal{L}(\boldsymbol R)
    =
    \left\|
        \widehat{\boldsymbol q}(\boldsymbol R)
        -
        \boldsymbol q
    \right\|_2^2.
    \label{eq_reconstruction_discrepancy}
\end{equation}
The objective of a reconstruction algorithm that uses a hard or soft outlier-detection method is to minimize this discrepancy, while the objective of the adversary is to maximize the discrepancy by injecting the most effective adversary ratings into the matrix~$\boldsymbol R$. 

We formulate the attack-generation procedure in a black-box setting. The attack is black-box with respect to the internal reconstruction and outlier-detection pipeline: the attacker has no access to its implementation, internal parameters, or derivatives. The attacker does, however, use the ground-truth quality vector in the experimental objective and receives query-based feedback in the form of the resulting discrepancy $\mathcal{L}$ to search for rating configurations that increase the reconstruction error.

Let $\boldsymbol R\in\mathbb{R}^{I\times J}$ denote the rating matrix
provided by legitimate subjects, where $I$ is the number of 
subjects and $J$ is the number of images. The adversary does not modify or
replace these ratings. Instead, it generates an additional matrix of ratings
$\boldsymbol A\in\mathbb{R}^{K\times J}$, where $K$ is the number of
attacking subjects. The rating matrix submitted to the
scale-reconstruction pipeline is the vertical concatenation
\begin{equation}
    \boldsymbol R^{\mathrm{aug}}(\boldsymbol A)
    =
    \begin{bmatrix}
        \boldsymbol R \\
        \boldsymbol A
    \end{bmatrix}
    \in\mathbb{R}^{(I+K)\times J}.
    \label{eq:augmented_rating_matrix}
\end{equation}
Thus, the attack consists of adding $K$ adversarial subjects to the original
set of $I$ subjects. The reconstructed quality vector and the corresponding
attack objective are evaluated as
\begin{equation}
    \widehat{\boldsymbol q}
    \left(
        \boldsymbol R^{\mathrm{aug}}(\boldsymbol A)
    \right)
\end{equation}
and
\begin{equation}
    \underset{\boldsymbol A}{\operatorname{maximize}}
    \quad
    \mathcal{L}
    \left(
        \boldsymbol R^{\mathrm{aug}}(\boldsymbol A)
    \right),
    \label{eq:attack_objective}
\end{equation}
respectively. Equivalently, the attack objective can be written explicitly as
\begin{equation}
    \underset{\boldsymbol A}{\operatorname{maximize}}
    \quad
    \left\|
        \widehat{\boldsymbol q}
        \left(
            \boldsymbol R^{\mathrm{aug}}(\boldsymbol A)
        \right)
        -
        \boldsymbol q
    \right\|_2^2.
    \label{eq:explicit_attack_objective}
\end{equation}
The original rating matrix $\boldsymbol R$ is fixed during optimization; only
the attacker matrix $\boldsymbol A$ is varied.

\subsection{Genetic algorithm for ACR attacks}
\label{sec:ga_acr}

For ACR, the attacker-rating matrix is discrete, and its entries satisfy
\begin{equation}
    A_{k,j}\in\{1,\ldots,5\},
    \qquad
    k=1,\ldots,K,\quad j=1,\ldots,J.
    \label{eq:acr_attack_constraints}
\end{equation}
Consequently, the attack search space contains $5^{KJ}$ possible matrices,
making exhaustive optimization impractical even for a moderate number
of attacking subjects and images. We therefore optimize the attack using a
basic genetic algorithm~(GA).

Each individual in the population represents a complete attacker-rating
matrix $\boldsymbol A$. The entries of an individual are encoded directly as
integer-valued genes, with one gene for each pair $(k,j)$. The fitness of an
individual is defined as the squared discrepancy obtained after augmenting
the legitimate rating matrix with the corresponding attacker matrix:
\begin{equation}
    f(\boldsymbol A)
    =
    \mathcal{L}
    \left(
        \boldsymbol R^{\mathrm{aug}}(\boldsymbol A)
    \right)
    =
    \left\|
        \widehat{\boldsymbol q}
        \left(
            \boldsymbol R^{\mathrm{aug}}(\boldsymbol A)
        \right)
        -
        \boldsymbol q
    \right\|_2^2.
    \label{eq:ga_fitness}
\end{equation}
The GA maximizes the fitness $f(\boldsymbol A)$.

The initial population consists of $P$ randomly generated individuals. Each
gene is sampled independently and uniformly from
$\{1,\ldots,5\}$. In every generation, individuals are ranked according
to their fitness. An elite subset containing the fittest $E$ individuals is
copied unchanged to the next generation. The remaining individuals are
generated by selecting the parent individuals, applying crossover, and then
applying mutation.

In each generation, the parents are determined using roulette-wheel
selection. The probability of selection of an individual is proportional to its
fitness, so individuals with larger squared discrepancies are more
likely to be selected.

Given two selected parent matrices, $\boldsymbol A^{(1)}$ and
$\boldsymbol A^{(2)}$, crossover is performed using a two-dimensional
row--column mask. Specifically, a random number of rows and a random number
of columns are selected to define the crossover mask. The corresponding
entries of the offspring are copied from one parent, while the remaining
entries are copied from the other parent. The complementary assignment is
used to generate the second offspring when two offspring are produced. In
this way, crossover exchanges groups of ratings associated with randomly
chosen subjects and images, rather than treating each rating independently.
Because both parents contain only valid ACR ratings, all offspring entries
remain in $\{1,\ldots,5\}$ after crossover.

Mutation is applied to the offspring according to the prescribed mutation
probability $p_{\mathrm m}$. When a gene is selected for mutation, its value is
replaced by another valid ACR rating in $\{1,\ldots,5\}$. The resulting
offspring are then evaluated using the fitness function in
\eqref{eq:ga_fitness}.

The GA uses a population size of $P$, an elite count of $E$, a crossover
probability of $p_{\mathrm c}$, a mutation probability of $p_{\mathrm m}$,
and a maximum of $G$ generations. All candidate individuals are evaluated
using the same fixed legitimate rating matrix, ground truth quality vector,
and scale-reconstruction procedure.

Because elitism preserves the best individual found so far, the best fitness
cannot decrease over generations. The adversarial matrix returned by the GA
is therefore the best individual in the final population,
\begin{equation}
    \boldsymbol A^\star_{\mathrm{GA}}
    =
    \underset{\boldsymbol A\in\mathcal{P}_G}
    {\operatorname{argmax}}
    \; f(\boldsymbol A),
\end{equation}
where $\mathcal{P}_G$ is the final GA population. This is the best solution found during the specified GA run and should be interpreted as an empirical attack result, not as a guarantee of the global optimum.

\subsection{Constrained continuous optimization for VAS attacks}
\label{sec:continuous_vas}

For VAS, attacker ratings are continuous variables. The attacker matrix
is therefore optimized subject to the nominal rating-scale constraints
\begin{equation}
    1\leq A_{k,j}\leq5,
    \qquad
    k=1,\ldots,K,\quad j=1,\ldots,J.
    \label{eq:vas_attack_constraints}
\end{equation}
The optimization problem is formulated as follows
\begin{equation}
    \underset{\boldsymbol A}{\operatorname{minimize}}
    \quad
    -
    \mathcal{L}
    \left(
        \boldsymbol R^{\mathrm{aug}}(\boldsymbol A)
    \right)
\end{equation}
subject to the bounds in~\eqref{eq:vas_attack_constraints}. Equivalently,
the objective can be written as
\begin{equation}
    \underset{\boldsymbol A}{\operatorname{minimize}}
    \quad
    -
    \left\|
        \widehat{\boldsymbol q}
        \left(
            \boldsymbol R^{\mathrm{aug}}(\boldsymbol A)
        \right)
        -
        \boldsymbol q
    \right\|_2^2.
    \label{eq:vas_attack_objective}
\end{equation}

\begin{table}[t]
\centering
\renewcommand{\arraystretch}{1.0}
    \begin{tabular}{lcccc}
        \toprule
          & \multicolumn{2}{c}{\textbf{Robustness}}   & \multicolumn{2}{c}{\textbf{Accuracy}} \\
          & \textbf{Mean RMSD} & \textbf{Rank} & \textbf{Mean RMSE} & \textbf{Rank} \\
        \midrule
         LPCC   & \textbf{0.0182}  & \textbf{1} & 0.2870 & 7 \\
         SUREAL & 0.0422 & 2 & 0.1240 & 2  \\
         ESQR   & 0.0455 & 3 & 0.1487 & 5  \\
         ZREC   & 0.0504 & 4  & \textbf{0.1227} & \textbf{1} \\
         CB     & 0.0758 & 5 & 0.1287 & 3  \\
         KB     & 0.1017 & 6 & 0.1429 & 4  \\
         RMLE   & 0.1139 & 7 & 0.1511 & 6  \\    
        \bottomrule
    \end{tabular}
    \caption{Spammer attacks on established outlier-detection methods. Robustness (RMSD) for this specific type of outliers is not indicative for the accuracy (RMSE) of the method.}
    \label{tab_spammers}
\end{table}

For scale-reconstruction methods employing differentiable soft
outlier-processing, this problem can be solved using a constrained nonlinear
optimizer. The initial point is a feasible attacker
matrix $\boldsymbol A^{(0)}$, whose entries are initialized randomly within
$[1,5]$. At each objective-function evaluation, the candidate matrix is
concatenated with the fixed legitimate rating matrix and processed by the
complete scale-reconstruction pipeline. 
The objective returned to the optimizer is the negative of the resulting squared discrepancy. Gradients are estimated numerically from query-based objective evaluations; consequently, the attack does not require explicit access to the pipeline’s internal derivatives or implementation.
The bounds in \eqref{eq:vas_attack_constraints} ensure that the adversarial VAS ratings generated remain feasible ratings from real subjects.

Continuous optimization is applied only to reconstruction methods whose
outlier-processing operations are differentiable. Hard rejection rules are
excluded from this procedure because their discrete decisions can make the
objective function non-differentiable or discontinuous. Such methods would
require a derivative-free black-box optimizer and are outside the scope of the
present VAS experiments.

As with the ACR attacks, the legitimate ratings and the ground truth quality
values remain fixed throughout the optimization. Only the $KJ$ entries of the
attacker matrix are optimized. If multiple initial points are used, the final
attack is selected based on the largest obtained value of
$    \mathcal{L}
    \left(
        \boldsymbol R^{\mathrm{aug}}(\boldsymbol A)
    \right).
    \label{eq:best_vas_attack}
$

\begin{figure}[t]
    \centering
    \includegraphics[width=\linewidth]{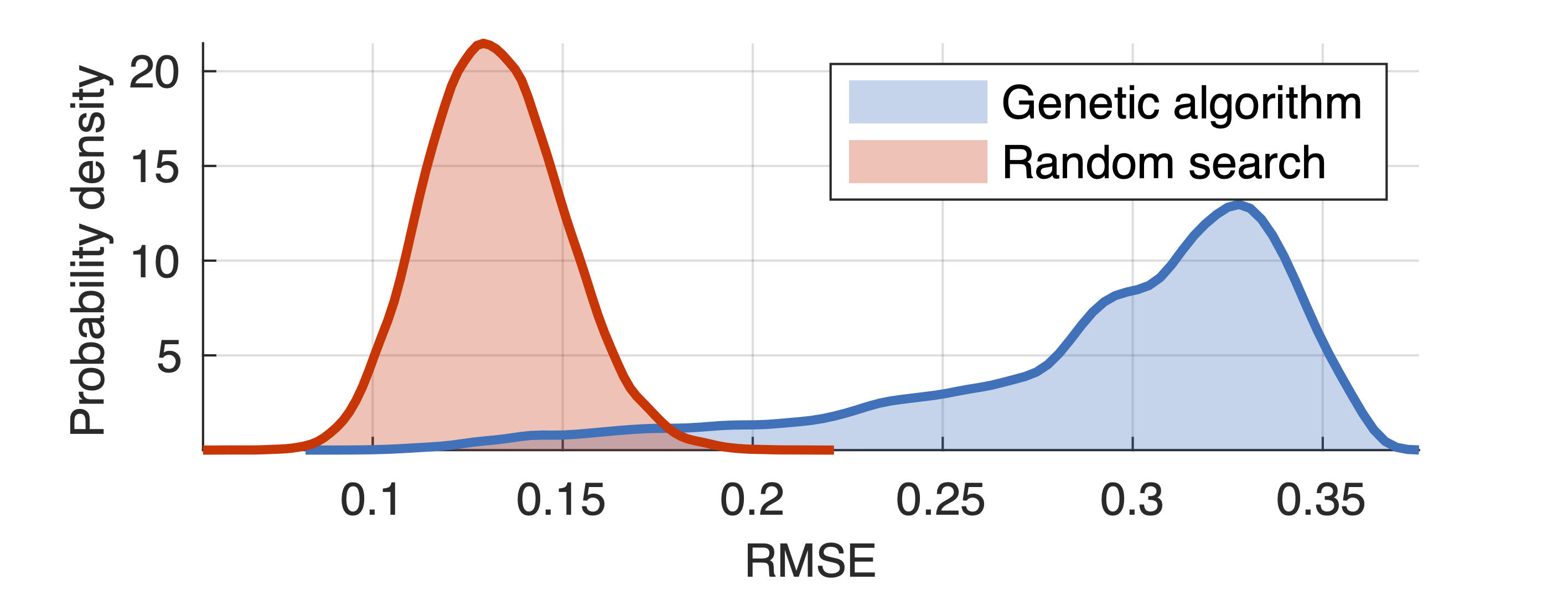}
    \caption{A single run of the GA attack on the kurtosis-based outlier detection (KB) produces 45,000 individual attacks (150 for each of 300 generations). The figure shows the density plot of the corresponding RMSE values, compared to 45,000 evaluations of randomly configured attacks.}
    \label{fig_ablation_random_search}
\end{figure}

\section{Results and discussion}
\label{sec_results}

The parameters for the simulated datasets were drawn from the KonIQ-10k dataset as explained in Subsection \ref{subsec:rating_simulation}. For each of our 250 simulated datasets, we randomly chose $I=30$ observers and $J=20$ images. All attacks were configured for $K=5$ attacking observers. These choices are similar to the cases considered in \cite{altieri2024subjective,tiotsop2022regularized,li2020simple,li2017recover}.

For convenience, we report the results for our experiments using the root-mean square error (RMSE) instead of the squared discrepancy of Eq.\ (\ref{eq_reconstruction_discrepancy}), i.e.,
\begin{equation}
\operatorname{RMSE}(\boldsymbol R)
=
\sqrt{
    \frac{1}{J}
    \sum_{j=1}^{J}
    \left(
        \widehat{q}_j(\boldsymbol R)-q_j
    \right)^2
}.
\end{equation}
where $J$ is the number of images.

Before showing results using the GA, we present an experiment that follows the procedures in the literature. It confirms that the most robust outlier method may, in fact, be the one with the worst quality reconstruction accuracy. We generated attacks from $K=5$ spammers who submitted random ACR ratings for each of the 250 simulated datasets. Table \ref{tab_spammers} shows that LPCC by far was the most robust outlier detection, yet the worst in terms of accuracy, again by a large margin. This can be explained by noticing that LPCC is too restrictive by wrongly rejecting many reliable observers as outliers.

\subsection{ACR attacks by the genetic algorithm}
\label{sec_results_ACR}

In principle, for the case of ACR, the genetic algorithm can be applied to optimize adversarial attacks for any reconstruction algorithm with an outlier removal component. However, in our experiments, we excluded several methods. RMLE was too time-consuming to run the genetic algorithm applied to all 250 datasets, requiring 11.25 million executions. DCC has the convergence issues mentioned before. Finally, \mbox{MAZsoft} and CBsoft were conceived after the main experiment only as derivatives of MAZ and CB to be used for VAS type ratings and continuous optimization. 

For each of the 250 dataset samples and all remaining outlier detection methods, the GA carried out the adversarial attack. For each method, we thus obtained 250 best attacks together with corresponding fitness values. 

We then carried out a small ablation study for just a single random simulated dataset, comparing the GA to a random search strategy for five spammers with an equal budget of fitness evaluations; see Fig.\  \ref{fig_ablation_random_search} for the case of the KB method. 
As expected, we found that the GA generated far better attacks.

\setlength{\tabcolsep}{3pt} 
\begin{table}[t]
\centering
    \begin{tabular}{lccccccc}
        \toprule       
         \textbf{Rank} & \textbf{Time} & \textbf{RMSE} &\textbf{RMSD} & \textbf{FPR} & \textbf{FNR} & \textbf{ACC}& \textbf{RAI} \\
        \midrule
        \phantom{1}1. HB & 69.3s & 0.206 & 0.106 & 0.050 & 0.298 &  0.915 & 0.050   \\
        \phantom{1}2. MAZ  & \phantom{2}2.2s & 0.212 & 0.119 & 0.051 & 0.482 &  0.888 & 0.078   \\
        \phantom{1}3. NLL  & \phantom{2}7.2s & 0.217 & 0.114 & 0.085 & 0.498 &  0.856 & 0.083  \\
        \phantom{1}4. ESQR  & 91.2s & 0.292 & 0.165 & \textbf{--} &  \textbf{--} & \textbf{--} & 0.123    \\
        \phantom{1}5. ZREC  & \phantom{2}3.2s & 0.350 & 0.312 & \textbf{--} &  \textbf{--} & \textbf{--} & 0.271   \\
        \phantom{1}6. SUREAL  & \phantom{2}7.5s & 0.351 & 0.321 & \textbf{--} &  \textbf{--} & \textbf{--} & 0.246   \\
        \phantom{1}7. NoOpt  & \textbf{--} & 0.372 & 0.358 & 0 & 1.000 &  0.857 & 0.143   \\
        \phantom{1}8. KB  & \phantom{2}5.0s & 0.373 & 0.306 & 0.022 & 1.000 &  0.838 & 0.146    \\
        \phantom{1}9. CB  & \phantom{2}8.9s & 0.432 & 0.344 & 0.230 & 0.973 &  0.664 & 0.175    \\
        10. LPCC  & 29.9s & 1.582 & 1.402 & 0.934 & 0.386 & 0.145 & 0.765   \\
        \bottomrule
    \end{tabular}
    \caption{Ranking of outlier-detection methods according to the worst-case reconstruction error (RMSE) under adversarial attack, averaged over 250 sample datasets. For the corresponding error distributions see Fig.\ \ref{fig_densityPlot_rmse}. The column RMSD shows the contradicting robustness measure, that compares reconstructions not with the ground truth but with those from the sample datasets. The run time is for all 300 generations of the GA for one sample dataset on an Apple M1 processor, averaged over ten trials. For the other columns see the text.
}
\label{tab_main}
\end{table}

\begin{figure}[t]
    \centering
    \includegraphics[width=\linewidth]{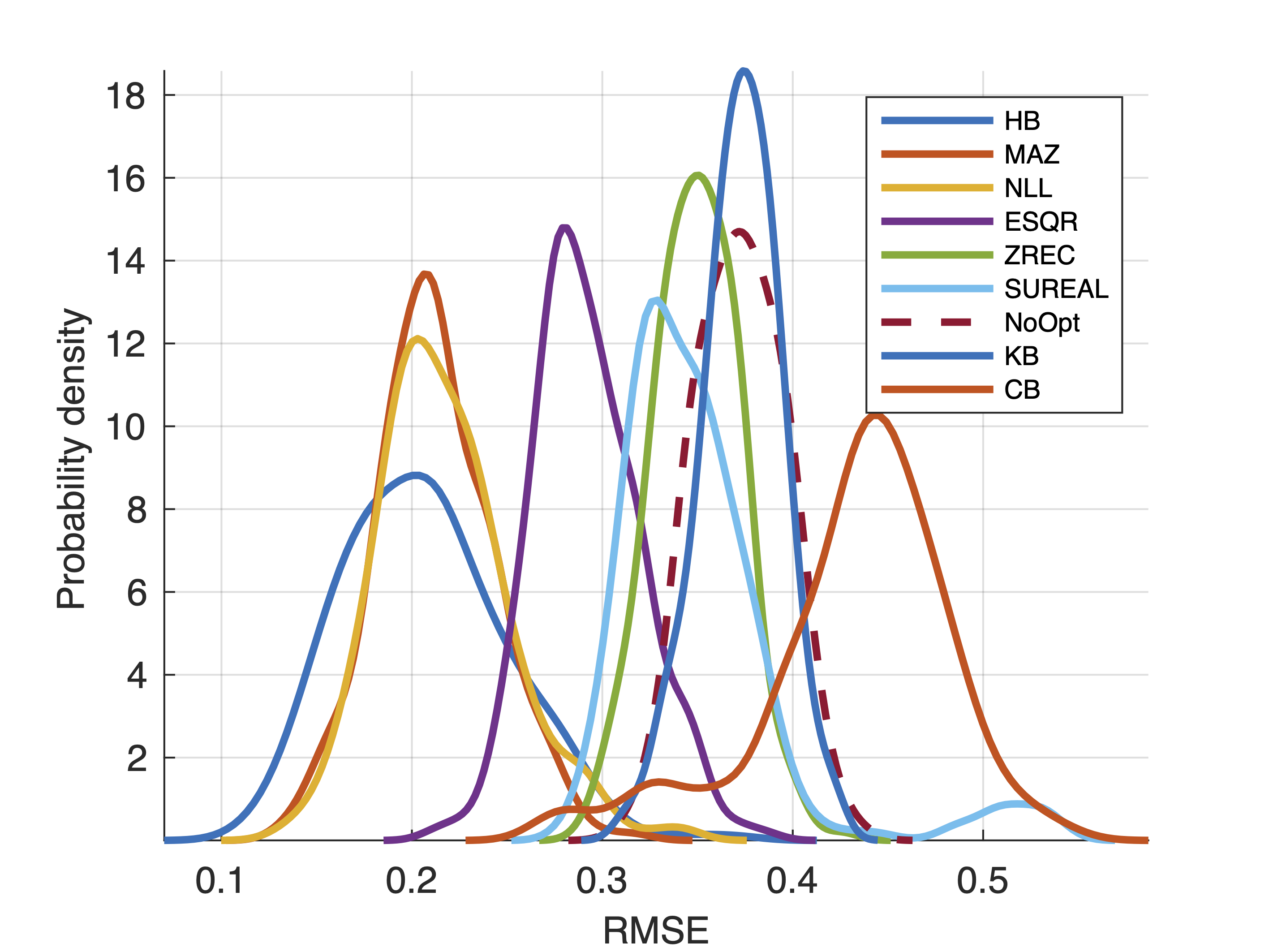}
    \caption{The curves are density plots of worst-case RMSE values for all outlier-detection methods (except LPCC) after adversarial attacks.}
    \label{fig_densityPlot_rmse}
\end{figure}

The result of applying the proposed empirical worst-case procedure is presented in Table \ref{tab_main} and Fig.\ \ref{fig_densityPlot_rmse}. The figure shows the densities for the RMSE distributions of the reconstructions for the 250 GA-attacked datasets. 
Under the specified GA attack procedure, the three hard outlier-detection methods HB (instructed to identify five outliers), MAZ, and NLL yielded the smallest mean RMSEs of about 0.2 which is about twice as large as the RMSE 0.115, obtained by the average MOS of the legitimate datasets without attack. All other methods have a mean RMSE of 0.29 to 0.44 with the exception of LPCC that was extremely vulnerable, having a worst-case mean RMSE of 1.8. 

For comparison, we included the trivial NoOpt method, which does not reject any observers as outliers. In this case, the best adversarial attack is easily identified. For each image, all adversarial ratings should be 1 or 5, depending on whether the MOS is above or below 3. 

Surprisingly, none of the hard outlier-detection methods of the ITU standards (KB, CB, LPCC\footnote{Poor performance of LPCC might be improved by reducing the correlation threshold. The standard ITU-T P.910 recommended 0.75 for the quality assessment of entertainment videos while our datasets are based on an image quality study, KonIQ-10k.}) were able to improve upon this trivial MOS of the attacked datasets by trying to detect and remove outliers.

For hard outlier detection, the mean false positive rate (FPR), mean false negative rate (FNR) and mean accuracy (ACC) of the classification are shown. FPR is the ratio of reliable observers that are regarded as outliers, and FNR is the ratio of undetected attackers. Furthermore, RAI is the mean remaining attacker influence, i.e., in the soft outlier detection the sum of the normalized weights of the attackers, and in the hard outlier detection the ratio of attackers in the inlier set. 

These numbers explain the lacking worst-case performance of LPCC and CB. For LPCC, 93\% of the reliable subjects were classified as outliers, and for CB, 97\% of the attackers were able to survive the outlier detection (on average). The RAI applies to all methods and provides a balanced view of FPR, FNR and ACC as it is highly correlated with the overall performance (RMSE), at least for the six best methods.

Some of our findings are expected. The recent state-of-the-art soft outlier-detection methods (SUREAL, ZREC, ESQR) rely on sophisticated subject models and weighting schemes that strike a balance by gracefully differentiating between observers that are more or less reliable in their ratings. Therefore, it is not a surprise that these have better worst-case performance leading to smaller MOS reconstruction errors in the presence of adversarial attacks, when compared to the more simple earlier hard outlier-detection methods (KB, CB, LPCC) in the ITU standards, followed by standard MOS computation.

However, it is surprising that the rather straightforward hard outlier-detection methods that had been unexplored (HB) or were newly introduced here (MAZ, NLL), together with simple MOS computation, exhibited worst-case performances that are significantly better by large margins than those of the soft outlier-detection methods employing complex subject models. However, it should be noted that for HB which achieved the best performance, we have chosen the true test-set contamination rate as the target number of subjects for outlier detection, $K=5$. This  can make the results overly optimistic.

\subsection{VAS attacks with a constrained continuous optimizer}
\label{sec_results_VAS}

\begin{table}[t]
\centering
\begin{tabular}{lccc}
\toprule
\textbf{Rank} & \textbf{Time} & \textbf{RMSE} & \textbf{RAI} \\
\midrule
1. MAZsoft & \phantom{2}0.4 & 0.265 & 0.176 \\      
2. ZREC & \phantom{2}1.1 & 0.360 & 0.521  \\
3. NoOpt & \textbf{--} & 0.367 & 0.143 \\
4. SUREAL & 10.3 & 0.372 & 0.980 \\
5. CBsoft & \phantom{2}0.5 & 0.385 & 0.156 \\
\bottomrule
\end{tabular}
\caption{Ranking of soft outlier-detection methods according to the worst-case reconstruction error (RMSE) under adversarial attack using a constrained nonlinear optimizer, averaged over 250 sample datasets. The corresponding error distributions are shown in Fig.\ \ref{fig_densityPlot_rmse}.  
}
\label{tab_main_soft}
\end{table}

For continuous VAS-type data, adversarial attacks using numerical gradient-based optimization are applicable only when the quality-reconstruction and outlier-detection pipeline is sufficiently smooth. This condition is not satisfied by hard outlier-detection methods: even an arbitrarily small change in an attacker’s rating can cause the subject to be included or removed from the reported inlier set, thus introducing a discontinuity in the objective function. Accordingly, this subsection considers only soft outlier-detection methods. ESQR is also excluded because its internal quantization of ratings for entropy estimation makes the resulting objective function non-differentiable. RMLE and DCC were excluded for the reasons given above. The methods considered for VAS attacks and optimization are therefore SUREAL, ZREC, and the adapted soft outlier-detection methods MAZsoft and CBsoft.

For adversarial attacks, we used MATLAB’s constrained nonlinear optimizer \texttt{fmincon} with its default \texttt{interior-point} algorithm. Gradients of the objective function were estimated using finite forward differences. With $K=5$ attacker subjects and $J=20$ stimuli, the attacker matrix $\boldsymbol A$ contains 100 optimization variables. Thus, estimating one forward-difference gradient requires 100 additional function evaluations. Convergence was assessed using a step tolerance of $10^{-6}$ for the rating variables. This value is the MATLAB default. With a much larger tolerance ($10^{-3}$) we did not obtain significantly different results, but substantially decreased the computational time for the iterative SUREAL method. The maximum number of iterations and function evaluations was set sufficiently high to allow convergence in all simulations, according to the criterion that the optimizer’s step size fell below the specified tolerance.

\begin{figure}[t]
    \centering
    \includegraphics[width=0.9\linewidth]{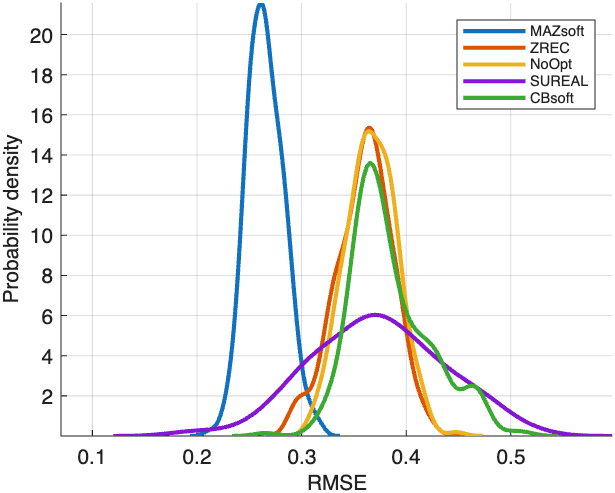}
    \caption{Density plots of worst-case RMSE values from 250 simulations each for soft outlier-detection methods after adversarial attacks using numerical optimization.}
    \label{fig_densityPlot_soft_rmse}
\end{figure}

For the VAS modality, we perform adversarial attacks using the same experimental protocol as for ACR. Specifically, 250 datasets, each comprising 30 subjects and 20 stimuli, were randomly generated, producing rating matrices $\boldsymbol{R}\in\mathbb{R}^{30\times20}$. For each dataset, a single optimization run was initialized with a randomly generated attacker matrix $\boldsymbol{A}\in\mathbb{R}^{5\times20}$.

Table \ref{tab_main_soft} and Figure \ref{fig_densityPlot_soft_rmse} summarize the performance of optimized VAS attacks, analogous to Table \ref{tab_main}, which reports the results of genetic-algorithm attacks for ACR. The false-positive and false-negative rates, as well as accuracy, are not applicable to soft outlier-detection methods. We therefore omit these measures and do not report the RMSD (Subsection \ref{SubSecRMSD}) again. The reported run times correspond to the optimization of one attack on one sample dataset using an Apple M3 processor, averaged over all 250 attacks.

First, the results show that the VAS attacks were successful, increasing the reconstruction errors, which are approximately 0.1 RMSE before the attacks. The ranking of the methods is similar to that observed for ACR. MAZsoft produced the smallest observed mean RMSE among the soft outlier-detection methods considered under the specified VAS attack procedure. It is based on mean absolute z-scores using a particularly simple algorithm and is also the fastest method. This finding is consistent with the position of MAZ near the top of the ranking for ACR with hard outlier detection.

The remaining tested methods performed similarly to the trivial NoOpt procedure that simply outputs the mean score of all subjects, including the attackers. ZREC and SUREAL gave average RMSE values close to each other and slightly larger than their corresponding ACR RMSE values by 0.1 and 0.2, respectively. The correlation-based method CBsoft provided the weakest resistance against adversarial attacks. 

The soft outlier-detection methods MAZsoft and CBsoft depend on the parameter $\alpha$, which determines the penalty applied to the ratings of subjects whose responses differ from consensus. We used the default value ($\alpha = 1$), selected based on visual inspection of several trial results.

Finally, the remaining attacker influence (RAI) reveals several noteworthy phenomena. In particular, the mean RAI for SUREAL is 0.980, substantially higher than the corresponding ACR value of 0.246. Thus, on average, the five adversarial subjects accounted for 98\% of the weights in the final weighted mean opinion score, while the total influence of the 30 legitimate subjects had decreased to only 2\%.

A closer examination of the 250 generated attacks shows that in 245 cases, one of the five adversarial subjects received a weight of 1.0, whereas all other subjects received zero weight. Consequently, the bias-corrected ratings of this single attacker completely determined the reconstructed quality values. These reconstructed values also represent the worst-case outcomes, as they maximize the discrepancy from the MOS obtained using only legitimate ratings. After reexamining the SUREAL results for ACR, we unexpectedly found that a few ACR attacks also caused a single adversarial subject to receive the entire weight.

A subject receives all of the weight when their ratings are assigned zero inconsistency. In turn, this implies that the reconstructed quality values are equal to the ratings of this subject, providing a fixed point of the iteration of the Alternating Projection solver in SUREAL, regardless of the ratings of all other subjects. The most successful attacks on SUREAL frequently produced solutions exhibiting this degenerate behavior. A more detailed analysis of the equations governing the iterations may help explain this phenomenon; however, such an analysis is beyond the scope of this paper and is left for future work.

For the correlation-based method CBsoft, the weights are distributed almost uniformly among legitimate and attacking subjects. The mean RAI is 0.156, which is close to the nominal fraction of attackers ($5/35 \approx 0.143$). The five attackers tend to converge to the same rating pattern in the 20 stimuli, consisting exclusively of ones and fives. For most, but not all, stimuli, the adversarial rating is 1 when the MOS of the original simulation data is below 3 and 5 otherwise. Thus, a subject who rates each stimulus as 1 or 5 according to whether its perceived quality is below or above fair quality may constitute an effective adversary. This finding suggests that CBsoft is not only the most vulnerable of the four soft outlier-detection methods tested under systematic adversarial attacks; in practice, a real subject could also exhibit this influential rating pattern, either intentionally or inadvertently.

For ZREC, the mean RAI is comparatively large, at 0.521. Consequently, each rating from an adversarial subject carries approximately seven times the weight of a rating from an original legitimate subject. The five attacking subjects tend to assign the same ratings to the stimuli as for CBsoft.

In general, continuous interval-scale ratings generally permit more severe adversarial attacks than ACR ratings. Injected attacker ratings can determine all reconstructed quality values, thereby dominating — and often effectively nullifying — the ratings provided by legitimate subjects. Evaluating both discrete ACR and continuous VAS-type outlier-detection methods under worst-case attacks allows one to identify promising techniques and characterize common rating patterns among influential outliers that evade detection with little or no penalty.

\section{Limitations and future work}
\label{sec_limitations}
The purpose of this study is to introduce and demonstrate an empirical worst-case framework, not to provide an exhaustive benchmark of datasets, attacker budgets, or optimization algorithms. The results should therefore be interpreted within the specific experimental setting described below.

The attack generators used here are proof-of-concept implementations for both rating modalities. A systematic comparison of attack-generation algorithms is outside the scope of this initial study. For example, simulated annealing, tabu search, particle swarm optimization, or differential evolution can be considered for ACR-attacks in place of the genetic algorithm. For continuous VAS data, we use a bounded local optimizer with numerical gradients. Its results may depend on initialization, finite-difference parameters, stopping criteria, and local optima.

For the simulated datasets, we selected 30 subjects and 20 items, consistent with the scale of typical laboratory studies, and introduced five attackers. The dependence of the estimated worst-case performance of the outlier-detection methods on these choices warrants further investigation.

In many cases, the most effective attack was obtained when all five attackers assigned the same rating to each image. In other cases, a single attacker completely determined the resulting MOS values. These observations suggest that it may be possible to restrict and simplify the attack model to a single attacker ($K=1$). Such a reduction would substantially accelerate both the genetic algorithm and the continuous optimizer. However, this simplification should be validated carefully, as multiple attackers may also produce collective effects that cannot be replicated by a single subject.

The ground-truth quality values were derived using the statistical subject model implemented in SUREAL. Although SUREAL represents a state-of-the-art approach, its model may still constitute a simplification of the underlying rating process. Analyzing and addressing its limitations remains an important challenge, both for improving scale reconstruction and for obtaining more reliable ground-truth estimates. Other statistical subject models could also be considered.

Ground-truth generation in this study was based exclusively on the KonIQ-10k dataset. Applying the proposed framework to other datasets containing individual-subject ratings may produce different results and could lead to different rankings of the outlier-detection methods.

The outlier-detection methods HB, NLL, MAZ, MAZsoft, and CBsoft depend on method-specific parameters: the target number of outliers for HB, thresholds for NLL and MAZ, and the parameter $\alpha$ for MAZsoft and CBsoft. In this study, these parameters were selected intuitively or on the basis of limited empirical evidence. A more systematic parameter-selection procedure could identify settings that improve the performance of these methods.

An interesting finding of the present study concerns the behavior observed in attacks against SUREAL. In most VAS attacks, and in a smaller number of ACR attacks, the alternating-projection procedure returned a spurious solution in which one adversarial subject had zero estimated inconsistency and consequently received all of the final weight. In such cases, the reconstructed quality values are determined entirely by that subject. This behavior produces highly effective attacks, however, its precise relationship with the iteration equations and initialization has not been investigated here. Future work should characterize the corresponding fixed points, assess their stability and frequency, and investigate whether regularization, alternative initialization, or constraints on the estimated subject weights can prevent such degenerate solutions without impairing normal operation.

A natural next step is to extend the framework to pairwise-comparison quality assessment. The evaluation principle remains unchanged: attacker responses would be selected to maximize reconstruction error relative to a model-based ground truth. Because pairwise responses are discrete, the attack would again form a combinatorial optimization problem. This extension would require a subject model tailored to pairwise comparisons, including an inconsistency parameter and an appropriate probabilistic reconstruction model. It would broaden the applicability of the proposed framework beyond single-stimulus ACR and VAS protocols.

Although worst-case analysis provides a rigorous stress test, it may also be informative to assess the accuracy of outlier-detection methods under more restricted and common forms of unreliability, such as occasional inattention among observers. This would extend the analysis of spammers presented in Table~\ref{tab_spammers}.

\section{Conclusions}
\label{sec_conclusions}

This work investigated the performance of outlier-detection methods under adversarial rating attacks in subjective visual quality assessment. We proposed an empirical worst-case framework in which attacker ratings are optimized to maximize the reconstruction error relative to a model-based ground truth. For discrete ACR data, the attack is formulated as a combinatorial optimization problem and addressed using a genetic algorithm. For continuous VAS data, we demonstrated a constrained nonlinear optimization approach for sufficiently smooth reconstruction pipelines. These procedures provide practical empirical stress tests. Because both search procedures are heuristic or local, the resulting values estimate the attainable worst-case error under the specified optimization procedures rather than proving the global worst-case error.

We also showed that the RMSD criterion commonly used in previous robustness studies can be misleading. In particular, a method may exhibit a small change between clean and attacked reconstructions, while still producing a large error relative to the ground truth. The LPCC results provide a striking example: LPCC achieved the smallest RMSD in the spammer experiment, but the largest reconstruction error. Consequently, evaluations based on clean-versus-attacked reconstruction differences should be complemented, whenever possible, by accuracy measurements with respect to an independent ground-truth model.

The tested outlier-detection methods differed substantially in their resistance to adversarial attacks. For ACR, the entropy-based method HB and the low-complexity hard methods MAZ and NLL achieved the best empirical worst-case performance, substantially outperforming the standardized methods SUREAL, KB, CB, and LPCC. For VAS, MAZsoft was the best-performing method and was also considerably faster than the other soft methods tested. 

These results suggest that simple outlier criteria can be highly competitive with, and in some cases more resistant than, sophisticated subject-model-based approaches. However, the performance of parameterized methods depends on the chosen thresholds and penalty parameters, which were not tuned systematically in this study.

The RAI analysis further revealed how the attacks succeed. For SUREAL, single adversarial subjects often obtained all  the final MOS weight. ZREC also assigned attackers substantially greater influence than legitimate observers. More generally, continuous VAS ratings allowed for more severe attacks than discrete ACR ratings, and coordinated attacker ratings could effectively dominate or even eliminate the contribution of legitimate observers.

The proposed framework, evaluated on 250 simulated datasets generated from SUREAL-based subject models fitted to KonIQ-10k, provides a reusable methodology for comparing outlier-detection and MOS-reconstruction methods under structured adversarial stress. The ACR and VAS experiments should be regarded as proof-of-concept instantiations of this methodology: they demonstrate the feasibility and diagnostic value of adversarial attacks, but do not exhaust the space of attack-generation algorithms, datasets, attacker budgets, or rating protocols. Extending the framework to pairwise comparisons and alternative attack optimizers is a natural direction for future work. We also make the implementation available to facilitate such extensions.\footnote{\url{github.com/UniStuttgart-VISUS/AdversarialAttacksForOutlierDetection}}

\noindent \textbf{Acknowledgment.}
The author thanks Tim Bleile for his contribution to our previous work \cite{saupe2025robustness} that was extended here. 

\bibliographystyle{IEEEbib}
\bibliography{ref}
\end{document}